\documentclass[twocolumn,prl,aps,showpacs]{revtex4}
\usepackage{amsmath}
\usepackage{epsfig}

\begin{document}

\title{Rosen-Zener interferometry with Ultracold Atoms}
\author{Li-Bin Fu$^1$}
\author{De-Fa Ye$^{1}$}
\author{Chaohong Lee$^2$}
\author{Weiping Zhang$^3$}
\author{Jie Liu$^1$}
\affiliation{$^1$Institute of Applied Physics and Computational
Mathematics, Beijing
100088, P. R. China\\
$^2$Nonlinear Physics Center and ARC Center of Excellence for
Quantum-Atom Optics, Research School of Physical Sciences and
Engineering, Australian
National University, Canberra ACT 0200, Australia\\
$^3$State Key Laboratory of Precision Spectroscopy, Department of
Physics, East China Normal University, Shanghai 200062, P. R. China}

\begin{abstract}
We propose a time-domain "interferometer" based on ultracold Bose
atoms loaded on a double well potential. By the adiabatic
Rosen-Zener process, the barrier between two wells is ramped down
slowly, held for a while, then ramped back. Starting with a coherent
state of double well system, the final occupations on one well show
interesting interference fringes in the time-domain. The fringe
pattern is sensitive to the initial state, the interatomic
interaction, and the external forces such as gravity which can
change the shape of the double well. In this sense, this
interferometric scheme has the potentials for precision measurements
with ultracold atoms. The underlying mechanism is revealed and
possible applications are discussed.
\end{abstract}

\pacs{03.75.-b, 05.45.-a, 03.75Kk, 42.50.Vk } \maketitle


Quantum interference is one of the most fundamental and challenging
principles in quantum mechanics, and has various applications in
high-precision measurement and quantum coherent control
\cite{a1,a2,a5}. Recently, a number of quantum interference
experiments have been performed with Bose-Einstein condensates
(BECs) \cite {CI-separated-BEC,CI-doublewell-BEC,CI-BEC-lattices},
where the coherent matter wave serves as phase coherent sources.
Using the device with matter waves instead of photons can improve
the measurement precision by a factor of $10^4 $ as shown in
\cite{22,23}, and the limit of phase sensitivity of standard
interferometer can be surpassed by matter wave interferometers \cite
{new1}. In particular, well-developed techniques in preparing and
manipulating BECs in the double well brought a new research surge
\cite {ket1,nt,rec,lgc,ag}. All these experiments demonstrated the
macroscopic quantum coherence of double well BECs with a spatial
interferometer.



In this letter, we propose an interferometer realized by the
adiabatic Rosen-Zener process with double-well BECs. The Rosen-Zener
model, proposed by Rosen and Zener to account for the spin-flip of
two-level atoms interacting with a rotating magnetic field in
Stern-Gerlach experiments \cite{RZT}. Here, in a double well scheme,
the adiabatic Rosen-Zener process is performed by slowly lowering
the barrier of the double well to a height and holding at this
height, then adiabatically lifting it back to the original height.
Due to inter-atom interaction, the system is an intrinsic nonlinear
system. Through such a nonlinear Rosen-Zener process, the system
shows marvelous interference effects in the time domain.


\begin{figure}[tbh]
\begin{center}
\rotatebox{0}{\resizebox *{8.0cm}{5cm} {\includegraphics
{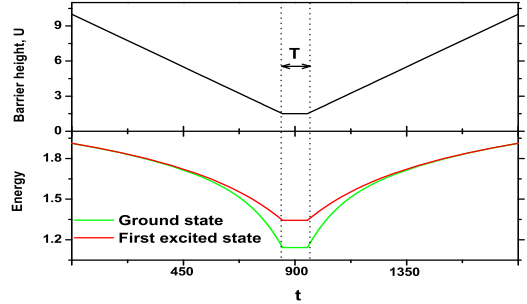}}}
\end{center}
\caption{Schematic diagram for an adiabatic Rosen Zener process. Upper
panel: the barrier height between two wells in time sequence. Bottom panel:
the energies of the Ground state and the first excited state in time with
varying barrier.}
\end{figure}

We consider a Bose atomic condensate trapped in a double well
potential with strongly transverse confinement. The dynamics obey a
1D model
\begin{equation}
i\hbar \frac{\partial }{\partial t}\Psi (x,t)=H_{0}\Psi (x,t)+\lambda |\Psi
(x,t)|^{2}\Psi (x,t),  \label{gpe}
\end{equation}
where $H_{0}=-\left( \hbar ^{2}/2m\right) \left( \partial
^{2}/\partial x^{2}\right) +V(x)$, $\lambda =8N\pi \hbar ^{2}\beta
_{1d}a_{s}/m$, $m$ is the single-atom mass and $a_{s}$ is the s-wave
scattering length describing the inter-atom interaction, $N$ is the
total particle number, and $\beta
_{1d} $ is the compensating coefficient for reducing transverse freedoms. $%
V(x)$ is a double well potential realized by superposing a Gaussian
barrier (see Fig. 1) on a harmonic trap
\begin{equation}
V(x)=\frac{1}{2}\omega ^{2}x^{2}+U\exp \left( -\frac{x^{2}}{2d^{2}}\right) ,
\label{trap}
\end{equation}
in which $\omega $ is the trapping frequency of the harmonic
potential, $d$ is the barrier width, and $U$ is the barrier height.
In the adiabatic Rosen-Zener process, we ramp the barrier height as
shown in Fig 1. The barrier is lowered slowly to a proper height,
and holding for time interval $T$, ramped up again slowly.

For a BEC trapped in a double well, when the barrier is high enough,
the ground state (GDS) $\left\vert \Psi _{g}\right\rangle $ and the
first excited state (FES) $\left\vert \Psi _{e}\right\rangle $ are
degenerate. The system has two local stable modes that are
superpositions of the ground state and the first excited state,
namely, $\left\vert \Psi _{L}\right\rangle =\frac{1}{\sqrt{2}}\left(
\left\vert \Psi
_{g}\right\rangle -\left\vert \Psi _{e}\right\rangle \right) $ and $%
\left\vert \Psi _{R}\right\rangle =\frac{1}{\sqrt{2}}\left(
\left\vert \Psi _{g}\right\rangle +\left\vert \Psi _{e}\right\rangle
\right) .$ For the two local stable modes, almost all the atoms are
localized in a single well, which are well-know as self-trapping
states \cite{smerzi,self,mil}.

\begin{figure}[tbh]
\begin{center}
\rotatebox{0}{\resizebox *{9.0cm}{5.5cm} {\includegraphics
{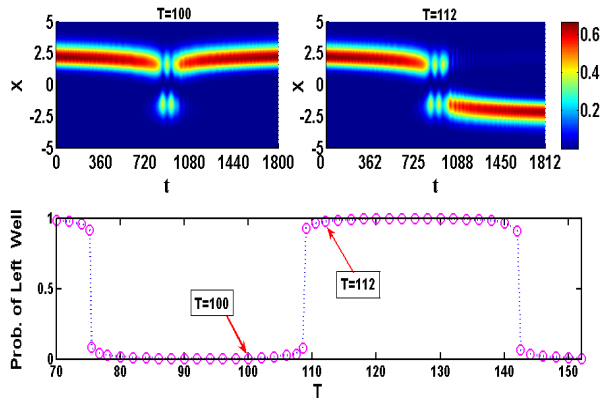}}}
\end{center}
\caption{Coherent transition of atoms by the Rosen Zener process.
The upper figures give two example of such a process. For the
holding time $T=100$ (left figure), atoms are still in the initial
well, however, for $T=112$ (right one) almost all atoms are
transferred to another well. The lower panel shows the probabilities
of occupation on the left well for different holding times $T$ (all
atoms in right well initially)} \label{rz}
\end{figure}

Throughout the entire process, the barrier is changed very slowly so
that the excitations to high eigenstates are very small. Hence, we
expect the atoms
still occupy only on the ground state and the first excited state at final, $%
\left| \Psi _{f}\right\rangle =c_{0}\left| \Psi _{g}\right\rangle
+c_{1}\left| \Psi _{e}\right\rangle .$ From the definition of the
local model, we have $\left| \Psi _{f}\right\rangle =a\left| \Psi
_{L}\right\rangle +b\left| \Psi _{R}\right\rangle $ with $a=\frac{1}{\sqrt{2}%
}(c_{0}-c_{1})$ and $b=\frac{1}{\sqrt{2}}(c_{0}+c_{1}),$ in which $%
|c_{0}|^{2}+|c_{1}|=1$. Therefore, the final probabilities on the
left and right well are
\begin{equation}
|a|^{2}=\frac{1}{2}-|c_{0}||c_{1}|\cos \theta ,\;|b|^{2}=\frac{1}{2}%
+|c_{0}||c_{1}|\cos \theta ,  \label{aa}
\end{equation}
where $\theta =\arg c_{1}-\arg c_{0}$ is the relative phase between
$\left| \Psi _{g}\right\rangle $ and $\left| \Psi _{e}\right\rangle
.$ The final occupations on each well are coherent interference of
the ground state and the first excited state, and depend on the
relative phase and probabilities of the two state. The final
occupation of one well severs as the interferometer ''output
ports''.

First we consider when the BECs are initially localized in one well
(e.g., the right one) and the barrier between two wells is
sufficiently high. The dynamics of BECs in the designed Rosen-Zener
process is demonstrated by Fig. 2. The results are obtained by
directly solving the GP equation using the operator-splitting
approach for the following dimensionless parameters: the trapping
frequency $\omega =0.2\pi $, the barrier width $d=1/\sqrt{2}$, the
initial and final height $U_{0}=U_{f}=10$, the lowest height
$U_{h}=1.5,$ and the ramping rate $\alpha =0.01$. The upper two
figures show the density evolution during the Rosen-Zener process
with the holding time $T=100$ and $T=112$, respectively. During the
holding time, the Josephson oscillations are clearly visible.
Finally, all atoms completely localize in one well, and which well
is occupied depends on the holding time. Moreover, the process is
robust, that is, there exist several intermittent windows of
complete transfer and blockade in the holding time (see the bottom
panel of Fig.2). The period of the rectangle functions is determined
by the atomic interaction and tends to infinity at certain
interactions as will be shown latter.

We assume the system only occupies the ground and first excited states
during the adiabatic process, i.e.,
\begin{equation}
\Psi (x,t)=c_{0}(t)\Psi _{g}(x,U)+c_{1}(t)\Psi _{e}(x,U)  \label{ccc}
\end{equation}
where $\Psi _{g}(x,U)$ and $\Psi _{e}(x,U)$ are the ground and first excited
states for GP equation (\ref{gpe}) with the barrier height $U$, which obey $%
E_{j}\Psi _{j}=H_{0}\Psi _{j}+\lambda \Psi _{j}^{3}$, where $E_{j}$
is the
chemical potential for $\Psi _{j}$ ($j=e,g)$. Defining $%
z=|c_{1}|^{2}-|c_{0}|^{2}$ and $\theta =\arg (c_{1})-\arg (c_{0})$,
we obtain the equations for $z$ and $\theta $ from GP equation
(\ref{gpe}):
$\dot{z}=-\frac{\partial H}{\partial \theta },\dot{\theta}=\frac{\partial H}{%
\partial z}$ with the classical Hamiltonian
\begin{equation}
H=\delta z+\frac{\beta }{2}z^{2}+\frac{\Lambda }{2}(1-z^{2})\cos 2\theta ,
\label{ch}
\end{equation}
in which $\delta =E_{e}-E_{g}+\frac{1}{2}(\gamma _{gg}-\gamma _{ee})$, $%
\beta =\frac{1}{2}(\gamma _{ee}+\gamma _{gg}-2\gamma _{eg})$,
$\Lambda =\gamma _{eg}$, and $\gamma _{ij}=\lambda \int \Psi
_{i}^{2}\Psi _{j}^{2}dx$ ($i,j=e,g$). In the above deductions, the
integrals with odd powers of $\Psi _{e}$ and $\Psi _{g}$ are nearly
zero and have been omitted\cite{s2m}. In the upper panel of Fig. 3,
we plot $Ee-E_{g}$, $\gamma _{ij}$ as functions of the barrier
height $U$.

\begin{figure}[tbh]
\begin{center}
\rotatebox{0}{\resizebox *{9.0cm}{6.0cm} {\includegraphics
{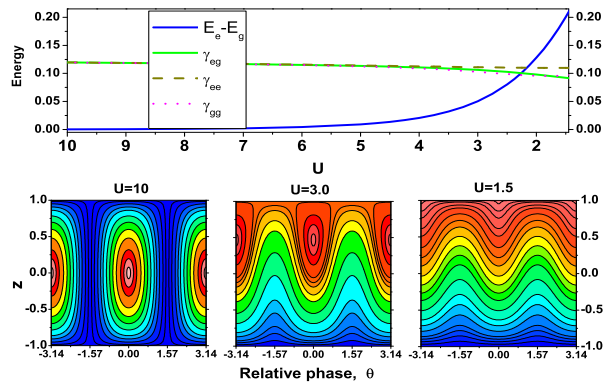}}}
\end{center}
\caption{The upper panel is for $\protect\gamma _{gg}$, $\protect\gamma %
_{eg} $, $\protect\gamma _{ee}$, and $E_{e}-E_{g}$ for different
barrier heights with the same parameters for Fig. 2. The bottom row
shows three phase spaces of the classical Hamiltonian (5) for
$U=10,3,1.5$ respectively.} \label{phase}
\end{figure}

The above classical Hamiltonian system has two axial fixed points at
$z=\pm 1$ independent of the relative phase, in which $z=-1$
corresponds to the ground state of GP equation (\ref{gpe}) and $z= +
1$ corresponds to the first excited state. The other fixed points of
the above classical
Hamiltonian can be obtained by solving the equation $\dot{z}=0$ and $\dot{%
\theta}=0$. For $\delta <(\Lambda -\beta )$, we obtain $(z^{\ast
},\theta ^{\ast })=(\frac{\delta }{\Lambda -\beta },0)$ and
$(z^{\ast },\theta ^{\ast })=(\frac{\delta }{\Lambda -\beta },\pi
)$. These two fixed points correspond to the self-trapping states in
double-well systems \cite {kiv,lbf}. When the barrier is low enough,
$\delta>(\Lambda -\beta )$, the two fixed points will merge into the
first excited states, i.e., $z=1$. In
bottom row of Fig. 3, we show the phase spaces of the classical Hamiltonian (%
\ref{ch}) for three typical values of barrier heights.

\begin{figure}[tbh]
\begin{center}
\rotatebox{0}{\resizebox *{8cm}{7.0cm} {\includegraphics
{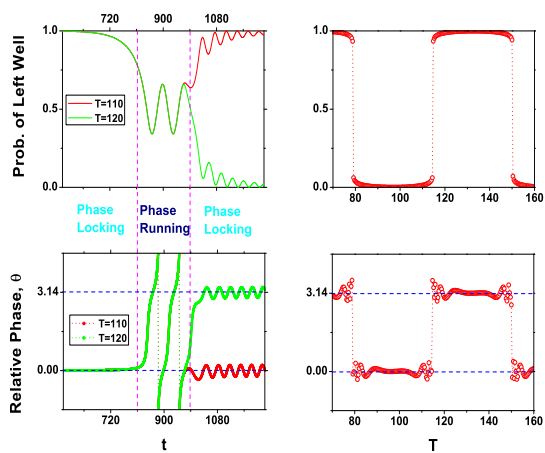}}}
\end{center}
\caption{Numerical results for the transfer probability and relative phase
obtained by the classical system (\ref{ch}). The left column shows the time
evolutions for $T=110,120$. The right column exhibits the dependence of
final occupations and phases on $T$. }
\label{class}
\end{figure}

\begin{figure}[tbh]
\begin{center}
\rotatebox{0}{\resizebox *{8.5cm}{5cm} {\includegraphics
{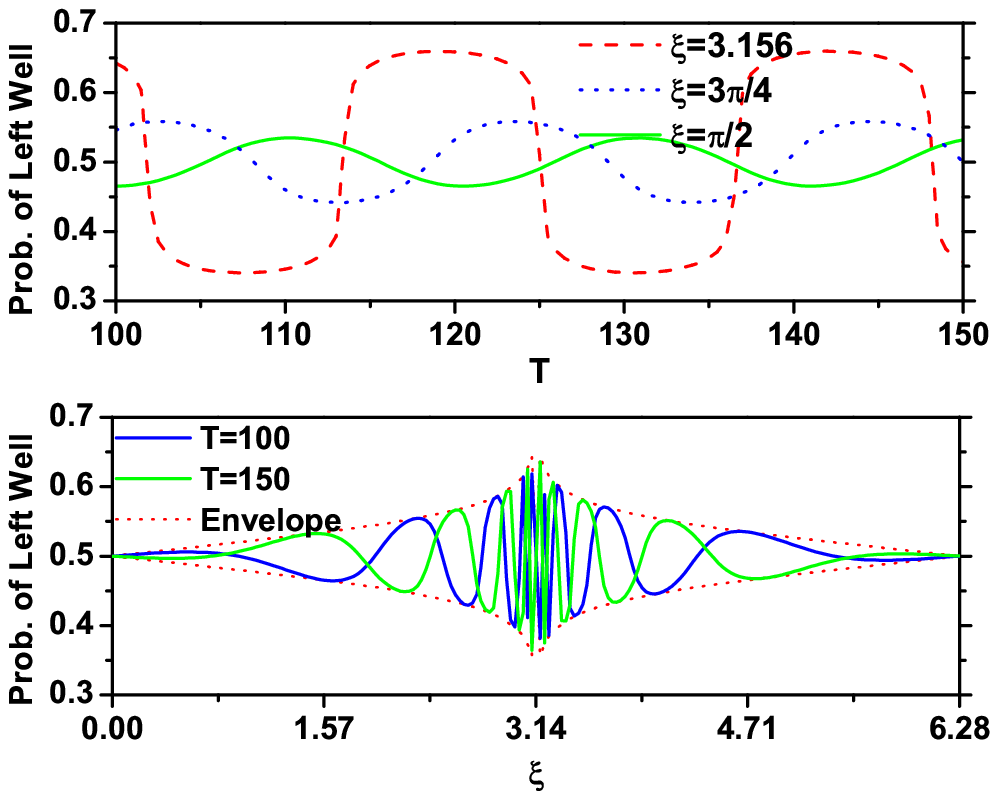}}} \rotatebox{0}{\resizebox *{8cm}{4.0cm}
{\includegraphics {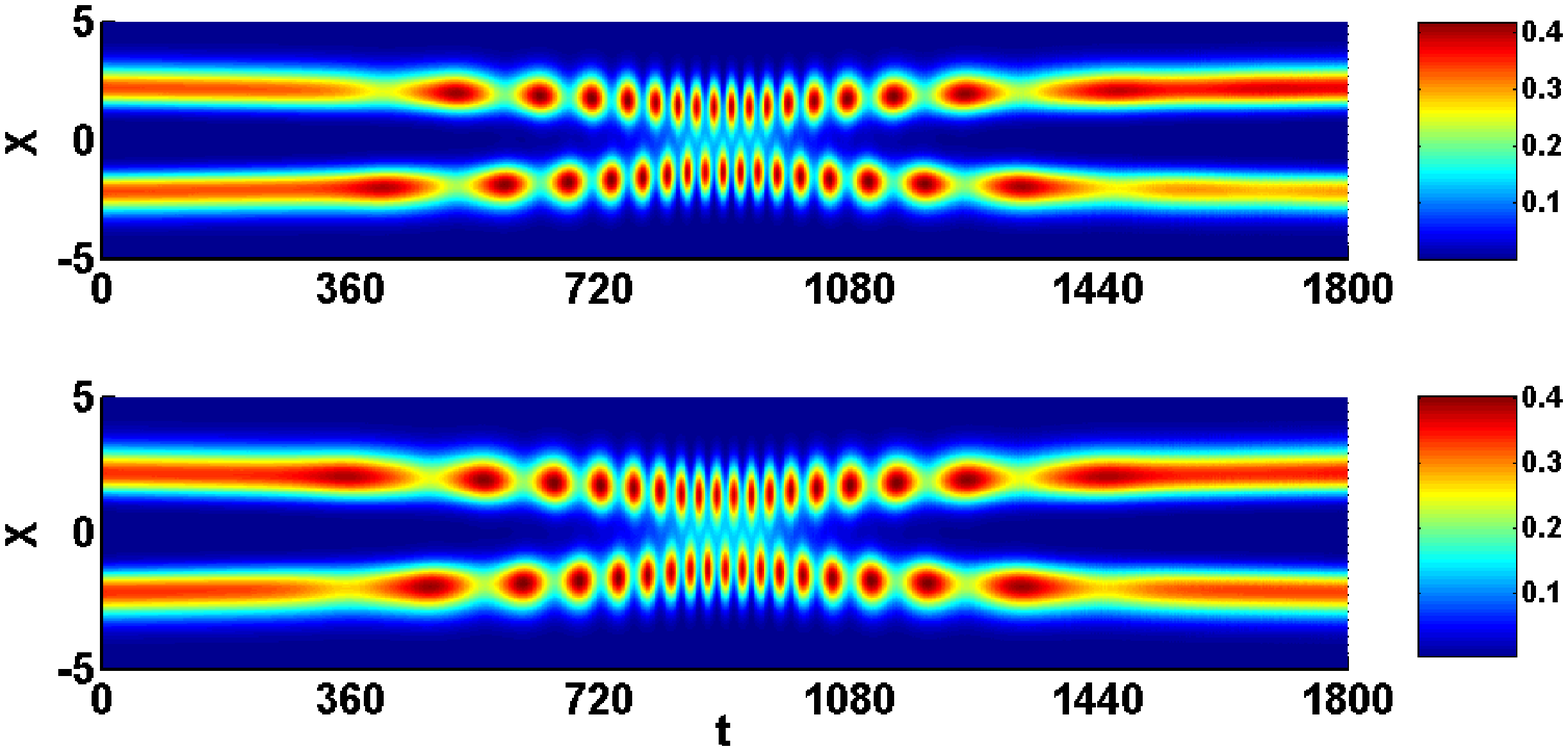}}}
\end{center}
\caption{The upper panel plots the final transfer probabilities of
different initial relative phase $\protect\xi $ vs the holding time
$T$ calculated with the classical model. The second row plots the
the final transfer probabilities vs different initial relative phase
$\protect\xi $ for $T=100$ and $150$. The dotted line (red) is the
envelope for the extreme values of every $\protect\xi $ when
changing the holding time $T$. The next two rows are two examples
obtained by directly solving the GP equation for $\protect\xi
=1.01\protect\pi $ and $1.02\protect\pi $, respectively.}
\label{ramsy}
\end{figure}

In the above process, the system is initially fully localized in the right
well, i.e., initially at the fixed point $(z^{\ast },\theta ^{\ast })=(0,0)$%
. As the barrier height decreases slowly, the system evolves along
the fixed point $(z^{\ast },\theta ^{\ast })=(\frac{\delta }{\Lambda
-\beta },0)$ until the barrier height is so low that $\delta
>\Lambda -\beta $. At this time, we hold the barrier unchanged for
time $T$. During the holding stage,
the state evolves close to the first excited state with relative phase $%
\theta $ running. The Josephson oscillation appears during this
stage. As the barrier height is raised again, the running phase
orbit will drop into phase locking orbit around one of the two fixed
points. Since the initial state is a fixed point, the adiabatic
evolution guarantees
the final state should be close to one of the two fixed points $(0,0)$ or $%
(0,\pi )$ \cite{adiab}. Because of the symmetry, the probability for
dropping into these two phase locking regions are the same, and the
period of the rectangular function is determined by the period of
the running phase orbit when the barrier is held. From the classical
model, the period can be calculated theoretically, which is
$\frac{2\pi }{\sqrt{(\delta -\beta )^{2}-\Lambda ^{2}}}$ (in which
the parameters are chosen as the values for the holding stage), and
for the above case it is about $35$, which is consistent with the
above calculation. The period can be controlled by the inter-atom
interaction.

In Fig. 4, using the classical Hamiltonian system (\ref{ch}) with
the barrier $U $ varying as showing in Fig. 1, and parameters
depending on $U$ from Fig. 3, we plot two typical probability
evolutions in the left column, and the final probabilities and
relative phase $\theta $ versus $T$ in the right column. These
figures show that the phenomena predicted by the GP equation can be
well reproduced and understood by the classical Hamiltonian system
(5).

Another interesting case is for the atoms populating evenly in two
wells (e.g., ground state). For such a case, the population
imbalance after the Rosen-Zener process is determined by both the
holding time $T$ and the initial phase difference $\xi $, the latter
can be controlled with the 'phase-imprinting' technique of shining
two laser beams with different
intersity\cite{phaseim}. 
Fig. 5, calculated by the Hamiltonian system (\ref{ch}), exhibits
the final populations in the left well versus the holding time $T$
for different $\xi $ (the top row), and the populations versus $\xi
$ for $T=100$ and $150$ (the second row). These figures show that,
after carrying out the Rosen-Zener scheme, the final population
occupations of the two wells depend on the relative phase $\xi $ as
well as the holding time $T$. In particular, for a fixed $T$, the
final occupations vary with the holding time $\xi $ showing a nice
interference pattern in the time domain. The interference pattern
depends on the nonlinear interaction and reduces to a sinusoidal
function in the absence of inter-atom interaction.

The numerical results also show that the final occupations are
sensitive to the phase $\xi$, especially around $\xi =\pi$. Thus
around the first excited state, the evolution is very sensitive to
the initial condition. These results are supported by directly
solving the GP equation. The bottom two rows of Fig. 5 are the
density evolutions for $\xi =1.01\pi $ and $1.03\pi$, respectively,
obtained by the GP equation, from which this sensitivity can be
found.

From the above simulations, we find that the final occupations on
the two wells sensitively depend on the initial conditions. Hence,
we can extract the initial information from the final occupations.
For example,
usually, the phase difference between the condensates in two wells
is measured by the spatial interference after withdrawing the
barrier. Here we show that the phase difference can also be measured
alternatively by interference in the time or phase domain. On the
other hand, with designing a Rosen-Zener scheme, one could realize
the double-well BECs with definite population imbalance and relative
phase serving as coherent matter wave source used for other
practical purpose.

Here, we only consider the case when the double well is symmetric.
If the double well is asymmetric, the interference fringe would be
sensitive to the energy bias, which affects the relative phase. The
gravitational field and any acceleration can create asymmetry in the
double well \cite{mel}, hence, the interference fringe obtained by
adiabatic Rosen-Zener interferometry can be used to measure the
gravitational field or any other acceleration.

In summary, a scheme for an interferometer with the matter wave in a
double-well potential serving as coherent sources is proposed. This
scheme is robust and realizable with present experimental
techniques. With it, the population imbalance of the atoms in two
wells shows interesting interference patterns in the time domain.
The fringe pattern is sensitive to the initial state, the
interatomic interaction, and the external forces such as gravity
which can change the shape of the double well. In this sense, this
interferometric scheme has the potentials for precision measurements
with ultracold atoms.

This work was supported by National Natural Science Foundation of
China (No.10725521,10604009, 10588402, 10474055), the National
Fundamental Research Programme of China under Grant No.
(2006CB921400, 2006CB921104, 2007CB814800), and the Australian
Research Council (ARC). WZ also acknowledges funding by the Science
and Technology Commission of Shanghai Municipality under Grants No.
06JC14026, the Research Fund for the Doctoral Program of Higher
Education No. 20040003101.

\end{document}